\documentclass[sigconf]{acmart}

\acmConference[EnCyCriS 2026]
{International Workshop on Engineering and Cybersecurity of Critical Systems}
{May 2026}
{Rio de Janeiro, Brazil}

\usepackage{enumitem}
 
\usepackage[english]{babel}
 \usepackage{comment}

\usepackage{amsmath}
\usepackage{graphicx}

\copyrightyear{2026}
\acmYear{2026}
\setcopyright{cc}
\setcctype{by}
\acmConference[EnCyCriS '26]{7th International Workshop on Engineering and Cybersecurity of Critical Systems }{April 12--18, 2026}{Rio de Janeiro, Brazil}
\acmBooktitle{7th International Workshop on Engineering and Cybersecurity of Critical Systems (EnCyCriS '26), April 12--18, 2026, Rio de Janeiro, Brazil}
\acmPrice{}
\acmDOI{10.1145/3786160.3788465}
\acmISBN{979-8-4007-2392-6/2026/04}

\title[A Transfer Learning Approach to Unveil the Role of Windows CCEs in IEC~62443 Compliance]{A Transfer Learning Approach to Unveil the Role of Windows Common Configuration Enumerations in IEC~62443 Compliance}

\author[Miguel Bicudo]{Miguel Bicudo}
\orcid{0000-0002-6266-4369}
\author[Estevão Rabello]{Estevão Rabello}
\orcid{0009-0002-7966-7723}
\author[Daniel  Menasché]{Daniel S. Menasché}
\orcid{0000-0002-8953-4003}
\affiliation{%
  \institution{Federal University of Rio de Janeiro}
  \city{Rio de Janeiro}
  \country{Brazil}
}

 \author{Paulo Segal}
 \orcid{0009-0000-3265-3047}
 \author{Claudio Segal}
 \orcid{0009-0000-4101-4191}
\affiliation{%
  \institution{Fluminense Federal University}
  \city{Rio de Janeiro}
  \country{Brazil}
}

\author{Anton Kocheturov}
\orcid{0000-0003-2549-9146}
\author{Priyanjan Sharma}
\orcid{0009-0001-7343-4876}
\affiliation{%
  \institution{Siemens Technology}
  \city{Princeton, NJ}
  \country{USA}
}

\begin{abstract}
Industrial control systems (ICS) depend on highly heterogeneous environments where Linux, proprietary real-time operating systems, and Windows coexist. 
Although the IEC~62443-3-3 standard provides a comprehensive framework for securing such systems, translating its requirements into concrete configuration checks remains challenging, especially for Windows platforms.   In this paper, we propose a transfer learning methodology that maps Windows Common Configuration Enumerations (CCEs) to IEC~62443-3-3 System Security Requirements by leveraging labeled Linux datasets. 
The resulting labeled dataset enables automated compliance checks, analysis of requirement prevalence, and identification of cross-platform similarities and divergences. 
Our results highlight the role of CCEs as a bridge between abstract standards and concrete configurations, advancing automation, traceability, and clarity in IEC~62443-3-3 compliance for Windows environments.
\end{abstract}
\keywords{Industrial Control Systems, IEC~62443-3-3, Transfer learning}

\ccsdesc[500]{Security and privacy~Systems security}

\begin{document}
\maketitle

\section{Introduction}

Industrial control systems (ICS) are increasingly exposed to cybersecurity threats due to their growing interconnection with corporate and public networks. 
To address these risks, the IEC~62443-3-3 standard  provides a comprehensive framework of system security requirements that cover both technical and organizational aspects of ICS protection~\cite{da2025safety,cindric2025mapping,leander2019applicability}. 
Despite its relevance, automating the process of assessing compliance with IEC~62443-3-3 remains a challenge, particularly in heterogeneous environments where operating systems such as Linux and Windows coexist. 
While Linux systems have been the focus of several studies and datasets, there is a scarcity of systematic approaches to mapping and validating IEC~62443-3-3 requirements in Windows environments.

 \textbf{Why Windows?}
In critical OT environments, heterogeneous systems commonly combine multiple operating systems: proprietary real-time OS from vendors such as Siemens and Rockwell for PLCs, Windows for PC-based components like HMIs and SCADA functions, and Linux for Industrial Edge devices. Despite this diversity, there is limited work on mapping and validating IEC~62443-3-3 requirements across these platforms, especially when they operate together.  This  gap   motivates our decision to begin by focusing on Windows systems.

\begin{figure}[t]
\centering
\includegraphics[width=\columnwidth]{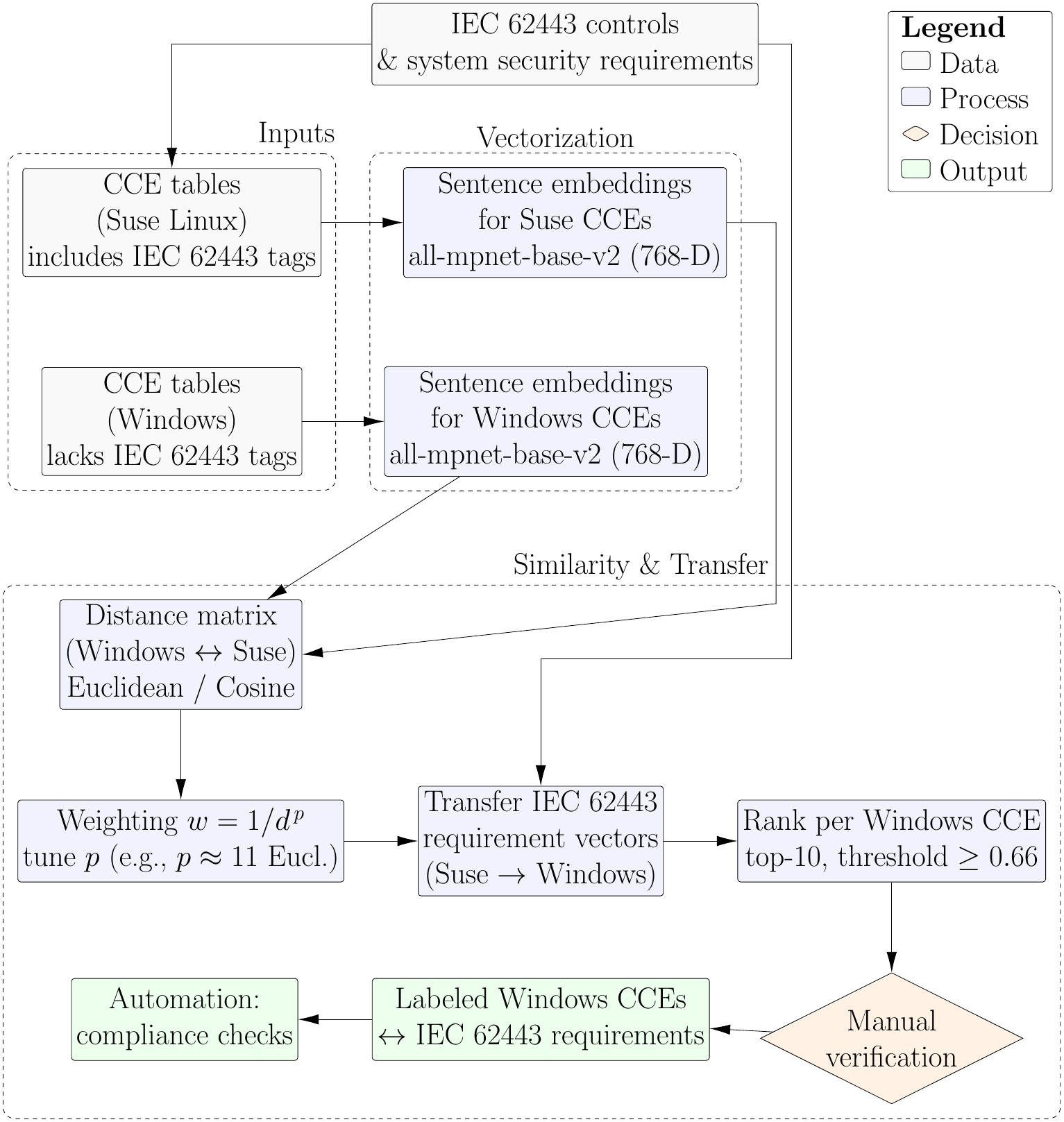}
\caption{Pipeline: Windows CCEs mapped to IEC~62443.}
\label{fig:iec62443_cce_pipeline_1}
\end{figure}

\textbf{Goals and contributions. }
Our goal is to automate the process of assessing compliance with respect to IEC~62443-3-3 in Windows systems. 
To this end, we make the following contributions:
 
 \textbf{Methodology to label a Common Configuration Enumeration (CCE)  dataset.} We propose a transfer learning approach to extend a labeled dataset of IEC~62443-3-3\cite{IEC62443-3-3} system security requirements originally related to Linux systems to Windows systems. Specifically, we map system security requirements from SUSE Linux configurations to their corresponding Windows configurations. The resulting mapping provides a rough approximation of system security requirements for Windows, which is then refined through manual inspection to validate and vet the results. The   methodology is summarized in Fig.~\ref{fig:iec62443_cce_pipeline_1}.
    
 \textbf{Analysis of the labeled CCE dataset.} Using the produced dataset, we analyze the prevalence of different system security requirements and identify clusters of requirements that tend to co-occur. Furthermore, we establish relationships between such clusters across Linux and Windows systems, highlighting similarities and divergences in their compliance landscapes.
   

Together, these contributions advance the state of the art in compliance assessment by providing both a methodology and empirical insights into how IEC~62443-3-3 requirements manifest in Windows environments. 
Our results aim to bridge the gap between abstract system security requirements and concrete system configurations, thereby supporting practitioners in automating compliance verification.

\section{Background and Initial Findings}

The Common Configuration Enumeration (CCE) is a collection of configurations uniquely identified by CCE IDs~\cite{niemi2024engaging,mann2008introduction}. The listing covers a range of software products, such as operating systems, web servers, databases, and browsers. An important aspect of this collection is that it is not a normalized database: each product has different sets of information organized in tables containing a CCE ID column and other columns with details about each configuration.  

A key product in this collection is the SUSE Linux operating system, which presents information about affected IEC~62443-3-3 system security requirements. Other products of interest, such as the Windows operating systems, do not provide the same information.

\textbf{IEC~62443-3-3. }  The IEC~62443-3-3 series of standards provides a comprehensive framework for securing industrial automation and control systems (IACS). Its layered approach emphasizes defense-in-depth, secure product development, and the clear definition of security levels across different zones and conduits. Despite this growing adoption, the integration of Common Configuration Enumerations (CCEs) within the IEC~62443-3-3 framework remains underexplored and not well understood. 


\textbf{Why mapping IEC~62443-3-3 SRs to CCEs and transferring knowledge?}
There is a clear need to align Windows Common Configuration Enumerations (CCEs) with IEC~62443-3-3 requirements to make configuration management consistent, auditable, and interoperable with other security frameworks. Mapping CCEs to specific IEC~62443-3-3 System Security Requirements (SRs) enables traceability, automation, and clarity in compliance verification, strengthening both compliance assurance and operational security.
To achieve this mapping despite the scarcity of labeled Windows data, we transfer knowledge from SUSE Linux configurations, for which IEC~62443-3-3 SR associations are available, to their corresponding Windows configurations. This transfer leverages the richer Linux-related dataset to infer likely Windows–requirement mappings, which are subsequently refined through manual inspection and validation.

\textbf{SUSE Linux CCEs and their related IEC~62443-3-3 system security requirements. }  
The CCE database does not require strict formatting within its datasets. Each product and each vendor may have different sets of data in their tables, along with CCE ID. In the case of SUSE Linux, the vendor provides descriptions of the CCEs and also relations with many security frameworks, including IEC~62443-3-3 system  security requirements.

\begin{figure}
    \centering
    \includegraphics[width=0.7\linewidth]{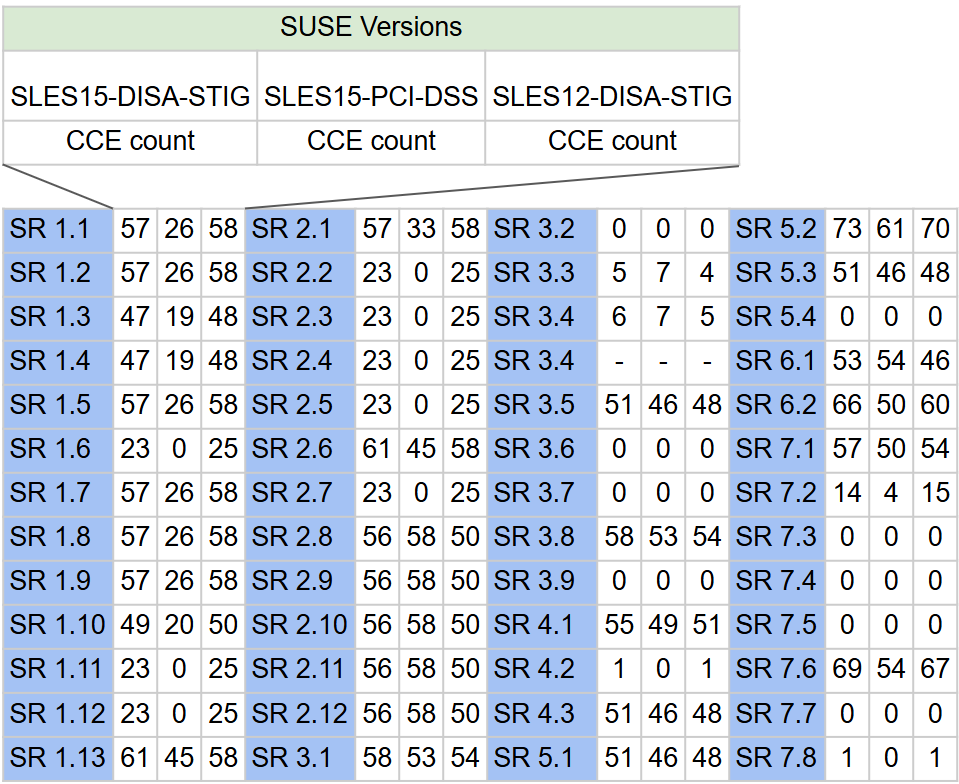}
    \caption{Counts of SRs and SUSE Linux configurations}
    \label{fig:sr-counts-suse}
\end{figure}

\textbf{Heterogeneity of references.}
To better understand the organization and prevalence of IEC~62443-3-3 SRs within SUSE Linux configurations, we analyzed the frequency of requirement mentions (Figure~\ref{fig:sr-counts-suse}). The results reveal substantial heterogeneity: some System  Security Requirements (SRs) 
 are referenced extensively, while others are rarely or never mentioned. This imbalance suggests that compliance with certain SRs may be more challenging to assess automatically using configuration-based evidence alone. Moreover, the frequent co-occurrence of particular SRs across multiple CCEs implies that some requirements tend to be evaluated jointly, forming meaningful groups of related controls.

\begin{figure}
    \centering
    \includegraphics[width=1\linewidth]{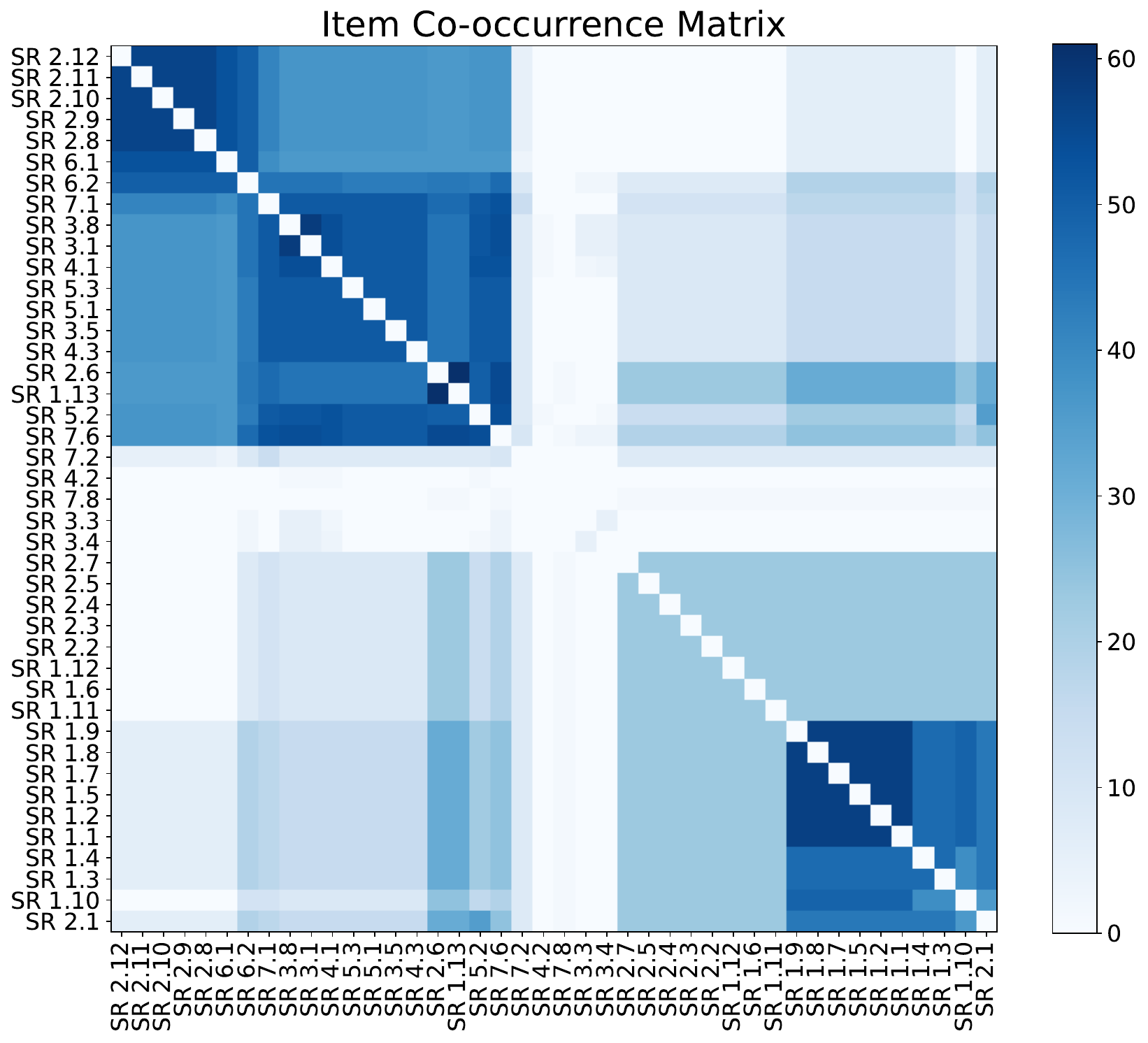}
    \caption{Co-occurrence matrix of SUSE Linux  System Security Requirements (SRs). We account for SUSE 15 DISA STIG. Each block corresponds to a set of SRs that tend to co-occur together. }
    \label{fig:cooccurrence-suse}
\end{figure}

This behavior is further illustrated in the sorted co-occurrence matrix in Figure~\ref{fig:cooccurrence-suse}, where clear block structures emerge. These blocks correspond to clusters of SRs that appear together in multiple configurations—indicating shared semantic or functional relationships. For example, some clusters concentrate around access control and authentication (e.g., SR~1.1–1.3), while others are dominated by system integrity and patch management requirements (e.g., SR~2.4–2.6). Such patterns reveal that security controls in practice are not applied independently but rather in groups that reflect common operational or architectural dependencies.

To explore this structure quantitatively, we applied unsupervised clustering to the SR co-occurrence matrix and used KeyBERT to extract representative keywords for each cluster. The resulting clusters and their representative terms are shown in Table~\ref{tab:clusters}, which provides a compact summary of how SRs organize into interpretable thematic groups. Each CCE was then labeled with the dominant cluster among its associated SRs, providing a higher-level abstraction of its security intent. For instance, clusters characterized by keywords such as “authentication,” “login,” and “access” capture configurations that strengthen user access management, while clusters emphasizing “update,” “integrity,” or “package” correspond to system hardening and maintenance controls. This taxonomy offers valuable insights into how different security functions are reflected across Linux configurations and supports the transfer of such knowledge to Windows environments.

\vspace{0.2in}
\noindent
\fbox{%
\parbox{\linewidth}{%
\textbf{Key findings based on Linux CCEs}

References to IEC~62443-3-3 System Security Requirements (SRs) are   heterogeneous: a few SRs dominate while many remain rarely referenced.
Frequent co-occurrences reveal clusters of SRs that represent coherent security themes (e.g., access control, integrity, patching).
These clusters provide a structured basis for transferring configuration-level security knowledge across platforms.
}%
}

\begin{table}[htbp]
\centering  
\caption{Clusters of SRs:   keywords and representative CCEs}
\begin{tabular}{p{1.5cm}|p{6cm}}
\textbf{Field} & \textbf{Value} \\ \hline
\textbf{Words} & auditd collects commands information privileged use ensure disk configure space \\
\textbf{CCEs} & CCE-85716-9, CCE-85587-4, CCE-85586-6, ... \\
\hline
\textbf{Words} & record events access attempts modify discretionary controls information ensure files \\
\textbf{CCEs} & CCE-83289-9, CCE-85610-4, CCE-85671-6, ... \\
\hline
\textbf{Words} & verify permissions disable files ssh del mount filesystems alt key \\
\textbf{CCEs} & CCE-85665-8, CCE-85625-2, CCE-85659-1, ... \\
\hline
\textbf{Words} & disable interfaces kernel set parameter icmp redirects password ipv4 accepting \\
\textbf{CCEs} & CCE-85669-0, CCE-85766-4, CCE-85715-1, ... \\
\end{tabular}
\label{tab:clusters}
\end{table}


\section{Methodology}

\subsection{Learning IEC~62443-3-3 SRs}

The association between Windows and SUSE Linux configurations is established through a distance-based weighting scheme. Each SUSE configuration is linked to a set of IEC~62443-3-3 requirements, which are transferred to nearby Windows configurations with weights that decrease as the distance increases, ensuring that closer configurations exert stronger influence in the mapping (see  Figure~\ref{fig:iec62443_cce_pipeline_1}).

\subsubsection{Embeddings, Distance Metric and Weighting}

The distance metric between CCEs is computed from vectorizations of each configuration. We first convert each CCE from its tabular format to JSON, representing each column as a dictionary. An embedding of this JSON object is obtained using the pretrained sentence embedding model \texttt{all-mpnet-base-v2}~\cite{reimers2019sentence}, one of the Sentence Transformers models from HuggingFace~\cite{wolf-etal-2020-transformers}. This model, based on MPNet~\cite{song2020mpnet} (a variation of BERT~\cite{devlin2019bert}), maps text into a 768-dimensional space where semantic similarity corresponds to geometric closeness.


After embeddings are obtained, we use the inverse of the Euclidean distance as the basis for transferring knowledge from SUSE to Windows configurations.  
For each Windows configuration $i \in \mathcal{W}$ and each SUSE configuration $j \in \mathcal{S}$, let $d_{ij}$ denote their Euclidean distance in the embedding space, and let $\mathbf{r}_j \in \{0,1\}^{R}$ represent the binary vector indicating which IEC~62443-3-3 SRs are associated with SUSE configuration $j$.  
The unnormalized score vector for Windows configuration $i$ is then defined as
\begin{equation}
    \tilde{\mathbf{s}}_i = \sum_{j \in \mathcal{S}} w_{ij}\, \mathbf{r}_j, 
    \qquad  w_{ij}=\frac{1}{d_{ij} + \varepsilon},
    \label{eq:unnormalized}
\end{equation}
where $\varepsilon > 0$ is a small constant to prevent division by zero.  
Finally, the normalized score vector $\mathbf{s}_i$, whose components $s_i{(r)} \in [0,1]$ represent the relative strength of the association between Windows configuration $i$ and requirement $r$, is obtained by rescaling all components to the interval $[0,1]$ using the $\ell_\infty$ norm, $
    \mathbf{s}_i = {\tilde{\mathbf{s}}_i}/{\|\tilde{\mathbf{s}}_i\|_{\infty}}.$ 
This normalization ensures that all inferred requirement scores are comparable across configurations, forming a consistent representation of IEC~62443-3-3 compliance likelihoods for Windows systems.

\subsubsection{Enhancing variability}
The naive approach of mapping SUSE to Windows CCEs yielded promising results but exhibited limited variability across the IEC~62443-3-3 SRs associated with Windows configurations. To enhance diversity in the resulting mappings, we applied a power-based transformation to the distance metric. This adjustment, conceptually related to the Box--Tidwell transformation~\cite{hastie2017generalized}, modifies the sensitivity of the metric by raising it to a power, thereby amplifying distinctions among nearby configurations and reducing the dominance of recurrent requirements. We let
$w_{ij}^{(p)} = w_{ij}^p$, and denote by $\tilde{\mathbf{s}}_i^{(p)}$ and $\mathbf{s}_i^{(p)}$ the corresponding unnormalized and normalized score vectors, respectively.

\paragraph{Selection rule (top-$K$ with threshold).}
For each Windows configuration $i \in \mathcal{W}$ and requirement index $r \in \{1,\dots,R\}$, 
let $\mathrm{rank}_i^{(p)}(r)$ denote the rank of $s_i^{(p)}(r)$ among 
$\{s_i^{(p)}(1), \dots, s_i^{(p)}(R)\}$ (with rank~1 corresponding to the highest score).  
Given a score threshold $\tau$ and a top-$K$ cutoff, the association between Windows CCE~$i$ 
and requirement~$r$ is retained if and only if $
   s_i^{(p)}(r) \ge \tau 
   \;\wedge\;
   \mathrm{rank}_i^{(p)}(r) \le K.$  
This rule ensures that only the most relevant requirements—according to both their relative 
ranking and absolute magnitude—are selected for each configuration.

To quantify how evenly the mapping covers different IEC~62443-3-3 requirements as the power parameter 
$p$ varies, we define the \textit{diversity index} $M(p)$. 
For each requirement $r$, we compute its maximum inferred score across all Windows configurations 
and then average these maxima over all requirements, $$M(p)   = \frac{1}{R} 
  \sum_{r=1}^{R} \max_{i \in \mathcal{W}} s_i^{(p)}(r).$$ 
A higher value of $M(p)$ indicates that more requirements achieve strong associations 
for at least one configuration, reflecting greater diversity in the transfer. 

\begin{figure} 
\centering
    \includegraphics[width=0.8\columnwidth]{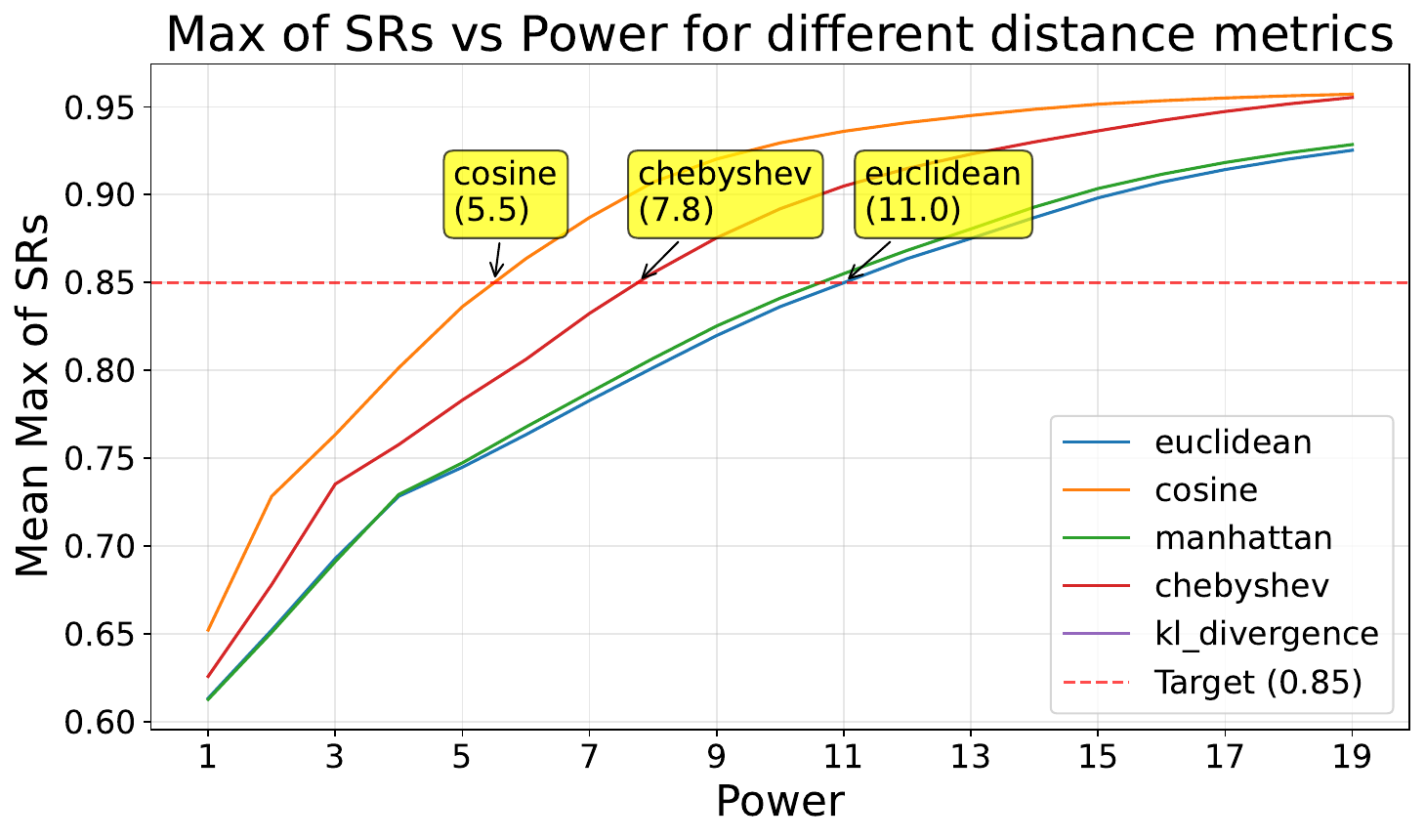} 
\vspace{-0.1in} \caption{Diversity index $M(p)$ as a function of $p$.} 
    \label{fig:embedding_distance_metrics_comparison}
\end{figure}

Figure~\ref{fig:embedding_distance_metrics_comparison} shows that the diversity index $M(p)$ increases monotonically with the power parameter $p$. To balance coverage and stability, we select $p$ such that $M(p) \approx 0.85$. As observed in Figure~\ref{fig:embedding_distance_metrics_comparison}, this value corresponds to $p = 5.5$ for cosine similarity.

\paragraph{Average list size.}
To quantify the combined effect of $p$, $\tau$, and $K$ on the selection outcome, we define 
the average list size as
\begin{equation}
  L(p) \;=\; \frac{1}{|\mathcal{W}|}
  \sum_{i \in \mathcal{W}} \sum_{r=1}^{R}
  1\!\left( s_i^{(p)}(r) \ge \tau \;\wedge\;
  \mathrm{rank}_i^{(p)}(r) \le K \right),
  \label{eq:avg-list-size}
\end{equation}
where   $1(\mathcal{C} )$ is the indicator  function of condition $ \mathcal{C}$.
This quantity represents the expected number of IEC~62443-3-3 requirements 
retained per Windows configuration under the chosen  parameters.

 \begin{table}[t]
\centering  \footnotesize
\caption{Summary of notation used in the mapping.} \vspace{-0.1in}
\begin{tabular}{p{1.8cm} p{6.2cm}}
\toprule
\textbf{Symbol} & \textbf{Description} \\ \midrule
$\mathcal{W}$ & Set of Windows CCEs (target domain) \\
$\mathcal{S}$ & Set of SUSE Linux CCEs (source domain) \\
$d_{ij}$ & Distance between Windows CCE~$i$ and SUSE  Linux CCE~$j$ in the embedding space \\
$\mathbf{r}_j$ & Binary vector of IEC~62443-3-3 reqs. for SUSE CCE~$j$ \\
$\mathbf{s}_i^{(p)}(r)$ & Normalized score vector,   $s_i^{(p)}(r) \in [0,1]$,  for IEC~62443-3-3 requirement $r$ under Windows CCE~$i$  \\
\bottomrule
\end{tabular} \vspace{-0.1in}
\label{tab:notation}
\end{table}
 \paragraph{Parameter tuning and manual verification.}
The parameters $p$, $\tau$, and $K$ were tuned empirically to balance coverage and precision. 
Unless stated otherwise, all results in this paper use cosine distance with $p = 5.5$ and $K = 10$. 
The threshold $\tau$ was chosen such that the average list size $L(p) = 5$, corresponding to $\tau = 0.68$. 
These settings provided a balanced trade-off between diversity and interpretability and are used consistently throughout the remainder of the paper. 
Using these parameters, we generated a table of Windows configurations and their inferred IEC~62443-3-3\cite{IEC62443-3-3} requirements. 
For each configuration, we listed the top 10 highest-ranked requirements that satisfied $s_i^{(r)}(p) \ge \tau$. 
Each mapping between a Windows configuration and an SR was then manually verified.

\subsection{Validation}

\textbf{Manual validation. }
The table produced in previous steps using the ML model  was verified by the two first authors.  The table comprises 3020 relations: each relation maps one of the 612 Windows configurations to at most 10 IEC~62443-3-3 SRs ($K=10$ in~\eqref{eq:avg-list-size}).  Each relation was assigned a label -- yes/no/maybe -- and an explanation for why a specific label was selected. Among the CCEs, 166 had no relationships with any SR. 
This is expected, as some SUSE Linux CCEs also have no associated SR. 


After manual filtering, we obtained 3020 analyzable CCE–SR relations (out of an initial 3060), of which 937 were accepted  with a Yes label. We were in doubt in 222 relations, marked with Maybe label, and declined 1861 SRs given by the ML model. The ratios of acceptance, doubts and declinations are 31\%, 7\% and 61\%. The missing 40 relations (1\% of them) were not analyzed due to problems in understanding the configuration descriptions (7 configurations), leaving us with 3020 relations.

 \textbf{Agreement against LLM. } 
To assess the consistency of our labeling, we compared each CCE–SR relation against Gemini~2.5~Flash. Overall, the manual labeling yielded 937 \emph{Yes}, 222 \emph{Maybe}, and 1861 \emph{No} associations, with 40 marked as \emph{N/A}. Gemini produced a similar distribution—953 \emph{Yes}, 196 \emph{Maybe}, and 1914 \emph{No}—indicating strong agreement in the global trends of inferred mappings across both annotation sources.

\begin{figure}[t]
  \centering
  \includegraphics[width=\linewidth]{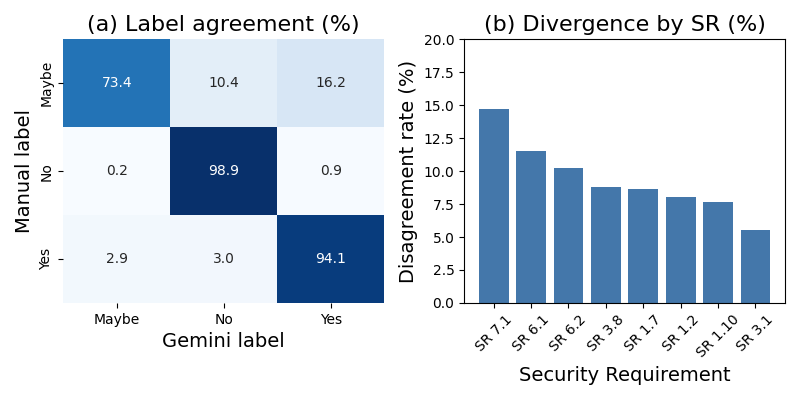} \vspace{-0.3in}
  \caption{
  (a) Confusion-matrix heatmap; 
  (b) Disagreement rates for  SRs with the highest divergence. } \vspace{-0.2in} 
  \label{fig:srplots}
\end{figure}

Gemini agreed with human assessors in \textbf{95.6\,\%} of all cases.
Agreement was highest for \texttt{No} (98.9\,\%) and \texttt{Yes} (94.1\,\%) labels, while most discrepancies stem from \texttt{Maybe} cases (73.4\,\%).
This pattern suggests that ambiguity in partial or conditional security relevance remains the main source of disagreement. Figure~\ref{fig:srplots} visualizes these results.



\section{Insights from Learned Mappings}\label{sec:insights}

This section analyzes the dataset of 446 Windows CCEs labeled against IEC~62443 SRs.  
Each configuration contains (i) textual descriptions of the setting, (ii) the underlying technical mechanism (e.g., registry keys, Group Policy Objects, or \texttt{auditpol} commands), and (iii) mappings to SRs assigned by both human annotators and Gemini~2.5~Flash.  
This analysis provides a detailed view of how IEC~62443 controls manifest across Windows configurations and how these mappings differ from their Linux counterparts.

The most frequently assigned SRs among the 937 accepted mappings, summarized in Table~\ref{tab:top-sr}, highlight the focus of Windows hardening practices.  
The distribution is dominated by SRs 7.6 (\textit{Security parameter enforcement}) and 5.2 (\textit{Network boundary protection}), followed by 6.2 (\textit{Continuous monitoring and logging}) and 6.1 (\textit{Audit log generation}).  
Together, these requirements emphasize that Windows configurations are primarily concerned with perimeter defense, audit enforcement, and the consistent application of security settings. 
In contrast, SRs focused on physical protection, external communications, or organizational policy were rarely mapped, underscoring the technical orientation of the dataset.

\textbf{Distribution and Thematic Concentration.}  
Nearly half of all accepted mappings correspond to SRs~5.2 and~7.6, showing that configuration-level compliance in Windows is highly driven by network and system-level enforcement mechanisms.  
This concentration indicates that automated compliance  is easier to achieve for operational controls that can be directly verified via local configuration parameters than for procedural  context-dependent SRs.  

\begin{figure}[t]
    \centering
    \includegraphics[width=0.7\linewidth]{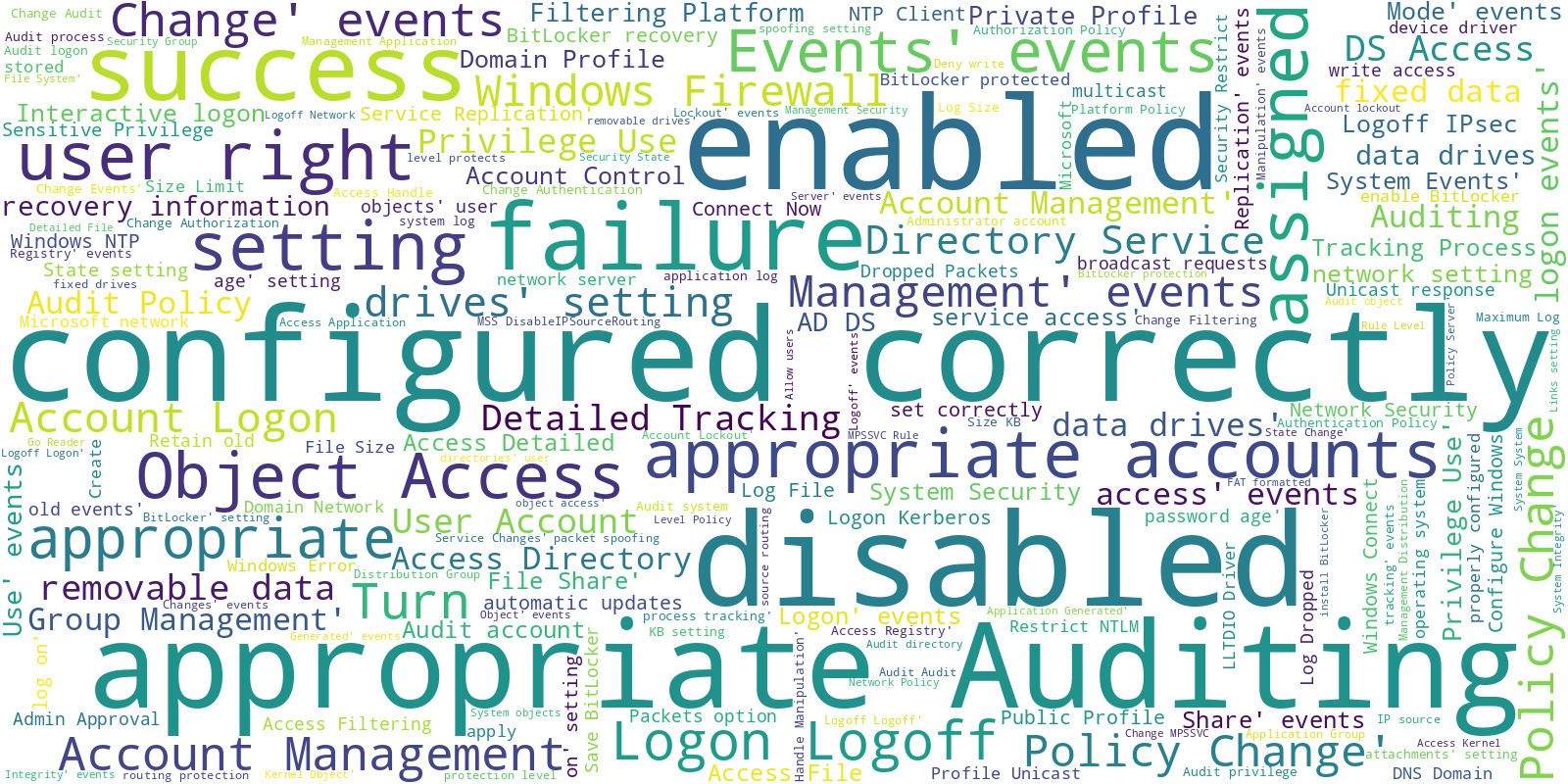} \vspace{-0.1in}
    \caption{Global word cloud of CCE descriptions.}
    \label{fig:wordcloud-global} \vspace{-0.2in}
\end{figure}

\textbf{Textual Patterns and Semantic Insights.}  
To characterize the textual layer of configurations, we analyzed the descriptions of all annotated Windows CCEs.  
Figure~\ref{fig:wordcloud-global} presents a global word cloud built from these texts, where token size is proportional to frequency across mappings.  
Terms such as \textit{configured}, \textit{enabled}, \textit{disabled}, \textit{policy}, \textit{enforced}, and \textit{auditing} dominate, illustrating the operational and directive nature of Windows configuration language.  
This vocabulary contrasts with the more procedural and descriptive tone observed in SUSE Linux CCEs (e.g., \textit{package}, \textit{service}, \textit{daemon}). 
These linguistic features also align with the prevalence of SR~7.6, as many descriptions   encode parameter validation or restriction mechanisms.

\begin{table}[t]
\centering  \footnotesize
\caption{Top Security Requirements (SR) by Frequency}
\vspace{-0.15in}
\begin{tabular}{l cc l}
\toprule
& \multicolumn{2}{c}{\textbf{Occurrences}} & \\
\cmidrule(lr){2-3}
\textbf{SR} & \textbf{All} & \textbf{Accepted} & \textbf{Main Theme} \\
\midrule
SR 5.2  & 400 & 155 & Network boundary protection \\
SR 7.6  & 332 & 282 & Security parameter enforcement \\
SR 6.2  & 245 & 147 & Continuous monitoring and logging \\
SR 2.9  & 218 & 4   & Session control \\
SR 2.12 & 212 & 8   & Credential robustness \\
SR 2.11 & 211 & 15  & Cryptographic key management \\
SR 2.8  & 203 & 19  & Event generation \\
SR 2.10 & 203 & 7   & Device identification \\
SR 6.1  & 174 & 142 & Audit log generation \\
SR 7.1  & 95  & 11  & Security function enforcement \\
SR 3.1  & 91  & 12  & Use control \\
SR 1.13 & 88  & 6   & Remote user authentication \\
SR 2.6  & 74  & 2   & Least privilege \\
SR 1.5  & 63  & 19  & Password policy enforcement \\
\bottomrule
\end{tabular}
\vspace{-0.3in}
\label{tab:top-sr}
\end{table}

\textbf{Cross-Domain Reflections.}  
Linux and Windows have some differences in their underlying security models: Linux relies on modular service configurations and file permissions, whereas Windows centralizes control via policies and registry-based constraints.  
From a transfer-learning standpoint, this asymmetry confirms that embedding-based similarity alone cannot fully capture cross-platform equivalences without semantic adaptation, highlighting the importance of manual validation in the final dataset.

\noindent
\fbox{%
\parbox{\linewidth}{%
\textbf{Key findings based on Windows CCEs}

Windows CCEs map to operational IEC~62443-3-3 requirements, highlighting the platform’s focus on access control and policy enforcement. The configuration language is prescriptive, dominated by terms like \textit{enabled} and \textit{auditing}, suggesting that compliance automation in Windows is well suited to technical controls.
}%
}

\noindent

\section{Conclusion}
This work tackled the challenge of operationalizing IEC~62443-3-3 compliance in Windows environments through the   use of CCEs. We introduced a transfer learning methodology that leverages   Linux datasets to infer IEC~62443-3-3 requirements for Windows configurations. The resulting labeled dataset enables automated compliance checks, facilitates the analysis of requirement prevalence, and uncovers   similarities and divergences between Linux and Windows compliance landscapes. By bridging the gap between abstract IEC~62443-3-3 requirements and concrete configuration practices, the approach enhances automation, traceability, and clarity in compliance assessment, paving the way for integration into continuous compliance pipelines and cross-framework alignment with standards such as NIST~800-82 and ISO/IEC~27001.

\textbf{Reproducibility. } The spreadsheet with Windows CCEs and  IEC~62443-3-3 mappings derived from  this work is available at \cite{Rabello2025WindowsCCE}.

\textbf{Acknowledgments. }  
This work was   partially supported by CAPES, CNPq, and FAPERJ under grant E-26/204.268/2024.

\bibliographystyle{ACM-Reference-Format}
\bibliography{main}

\end{document}